\documentclass{styles/svproc}
\usepackage{url}
\usepackage{eso-pic}

\usepackage{cite}
\usepackage[colorlinks = true,
            linkcolor = blue,
            urlcolor  = blue,
            citecolor = blue,
            anchorcolor = blue]{hyperref}
\usepackage{amsmath,amssymb,amsfonts}
\usepackage{algorithmic}
\usepackage{graphicx}
\usepackage{textcomp}
\usepackage{xcolor}
\usepackage{booktabs}
\usepackage{tabularx}
\usepackage{graphicx}
\usepackage{multirow}
\begin{document}
% \AddToShipoutPictureFG*{%
%   \AtPageUpperLeft{%
%     \raisebox{-1.5cm}{\hspace{1cm}%
%       \smallSpecial Session Invited Paper: Intelligent Machine Learning Frameworks for Biomedical and Bioinformatics Applications
%     }%
%   }%
% }
\mainmatter              % start of a contribution
\title{Adversarial Efficiency Degradation for Vision Transformer by Exploiting Input-adaptive Optimization: A Survey\\}
\titlerunning{Adversarial Efficiency Degradation}  % abbreviated title (for running head)
\author{Anadi Goyal\inst{1} \and Nandish Chattopadhyay\inst{1}
Anupam Chattopadhyay\inst{2} \and Chandan Karfa\inst{1}}
\authorrunning{Anadi Goyal et al.} % abbreviated author list (for running head)
% \tocauthor{Anadi Goyal, Nandish Chattopadhyay, Anupam Chattopadhyay, and Chandan Karfa}
%
\institute{Indian Institute of Technology Guwahati, Guwahati 781039, Assam, India,\\
\email{anadigoyal@rnd.iitg.ac.in},\\
\and
Nanyang Technological University, 50 Nanyang Ave, Singapore 639798}

\maketitle              % typeset the title of the contribution
\begin{abstract}
Vision Transformers (ViTs) increasingly rely on input-adaptive inference, such as token pruning and early halting, to meet energy and latency budgets. This survey examines a recent class of \emph{adversarial efficiency degradation attacks} that target those mechanisms to increase computation without necessarily degrading accuracy. We unify and compare two representative attacks: \emph{SlowFormer} (a universal adversarial patch) and \emph{DeSparsify} (per-image $\ell_\infty$ perturbations), across three popular token-pruning frameworks (A-ViT, ATS, AdaViT). We standardize reporting using GFLOPs, accuracy loss, and an \emph{Attack Success} (AS) measures how much of the model’s compute savings the attack takes away. Understanding these attacks is crucial for designing countermeasures that not only mitigate risk but also remain lightweight, since deployment often occurs in low-power settings such as mobile or embedded devices. To organize our analysis, we focus on three questions: (i) how input-adaptive optimizations (e.g., token pruning, early halting) create attack surfaces for efficiency degradation; (ii) how such attacks operate in practice and which optimizations are most vulnerable; and (iii) which defenses exist today and whether they meaningfully restore efficiency under attack.
\end{abstract}
% \begin{IEEEkeywords}
% Vision Transformers, Adversarial Efficiency Degradation Attacks and Token Pruning.
% \end{IEEEkeywords}

\section{Introduction} \label{sec:1}
 Transformers \cite{vaswani2017attention}, built on the self-attention mechanism, have demonstrated remarkable success in natural language processing (NLP) tasks such as machine translation, language modeling, and semantic analysis. Their architecture has also been adapted for computer vision (CV), leading to the introduction of Vision Transformers (ViTs) \cite{dosovitskiy2020image}. ViTs are now widely used in domains including image classification, object detection, segmentation, and video analysis. 
 
 Despite these advances, ViTs face significant efficiency challenges in resource-limited settings, as their computational cost grows quadratically with the number of tokens \cite{dosovitskiy2020image}. To address this, researchers have developed a range of techniques aimed at improving their efficiency. These approaches can be broadly grouped into two categories: \textbf{(1) \textbf{Static Methods}} that reduce computation uniformly, regardless of the input, and \textbf{(2) \textbf{Dynamic Methods}} that adjust computation dynamically depending on the input. Common strategies such as weight pruning or model quantization fall into the first category, where the reduction is fixed across all cases. However, in many real-world applications, the complexity of the input varies, and dynamic methods can save significant resources. For example, in facial recognition systems, analyzing a clear, well-lit face against a plain background requires much less computation than processing an image with multiple overlapping faces.
Such input-adaptive optimizations are particularly important in critical, resource-constrained environments. One such adaptive method is Token Pruning, in which tokens are dynamically removed based on their importance. Token Pruning is a prominent technique used for reducing computations in ViTs. Some popular token pruning approaches proposed include:
 ATS \cite{ats}, AdaViT \cite{adavit}, and A-ViT \cite{avit}, each of which adaptively removes uniformative tokens based on the
 complexity of the input image (i.e., input-dependent inference), resulting in improved throughput with a slight drop in accuracy.
\begin{figure}
    \centering
    \includegraphics[width=1\linewidth]{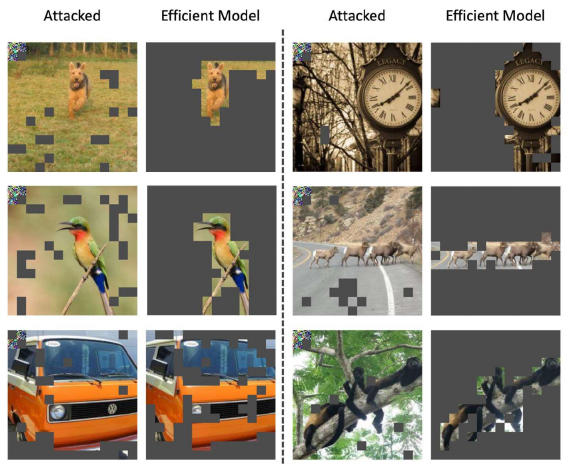}
    \caption{A-ViT token pruning on ViT-Small: \ SlowFormer patch \cite{slowformer} vs Clean sample. 
Grey marks pruned tokens. The adversarial patch causes fewer tokens to be removed than in the clean case, increasing retained tokens and thus increasing GFLOPs and power.}
    \label{fig:placeholder1}
\end{figure}

 While adaptability improves efficiency in many contexts, it also opens the door to new forms of security risks. Recent research highlights that adversaries can exploit these adaptive mechanisms like token pruning to mount adversarial efficiency degradation attacks on ViTs, where specially designed inputs force the model to perform unnecessary or excessive computations. Unlike conventional adversarial attacks, which typically aim to mislead the model into producing incorrect outputs \cite{cod_1}, these attacks target the computational process itself. The result is a sharp rise in inference-time costs, leading to longer latency and higher energy demands. Beyond wasted resources, the broader consequences are substantial: real-time healthcare monitoring systems may fail to deliver timely results, autonomous drones could experience faster battery draining during critical missions, and large-scale data centers might suffer from degraded throughput and escalating infrastructure costs. These attacks can take the form of a single-image attack \cite{cod_2}, \cite{cod_3}, where perturbations are crafted specifically for each input, or a universal adversarial patch, which, when pasted to any image, forces the model to perform additional computation and thereby increases power consumption \cite{cod_4}, \cite{cod_5}.
Figure \ref{fig:placeholder1} illustrates the tokens pruned (shown in grey) by the A-ViT \cite{avit} pruning method on the ViT-Small model, both in the absence of an attack and under the SlowFormer patch attack \cite{slowformer}. The adversarially perturbed image, containing a small patch, significantly reduced the number of tokens eliminated compared to the clean input. This reduction directly leads to higher computation and greater power consumption. Notably, despite the patch being relatively small and positioned in the corner of the image, it exerts an influence over the entire computational process of the network.

A new category of adversarial attacks, targeting efficiency rather than accuracy, remains largely unexplored in a structured manner. Understanding these attacks is crucial for designing effective countermeasures that not only mitigate the risks but also remain lightweight, since their deployment often occurs in low-power environments such as mobile or embedded devices.
To address this gap, the present article aims to \textbf{survey and systematize research on Adversarial Efficiency Degradation Attacks against ViTs}. 

To organize our analysis, we focus on the following key research questions:

\begin{itemize}
    \item \textbf{RQ1:} How do input-adaptive inference methods open attack surfaces that increase computation or energy?
    \item \textbf{RQ2:} What mechanisms enable efficiency-targeted attacks on ViTs to fool inference optimizations toward higher compute?
    \item \textbf{RQ3:} Which defenses keep attacks from raising compute yet preserve energy and latency advantages, and how robust are they?
\end{itemize}
Specifically, we examine the dynamic inference behaviors that create such vulnerabilities, review and explain the mechanisms of attacks proposed thus far, evaluate existing defense strategies, and identify key challenges along with directions for future research.
\section{Background and Related Work}
\subsection{Vision Transformers}
ViTs \cite{dosovitskiy2020image} employ a backbone of $L$ transformer encoder blocks, each composed of a Multi-Head Self-Attention (MSA) layer and a Feed-Forward Network (FFN). An input image $x \in \mathbb{R}^{C \times H \times W}$ is partitioned into $N$ patches, linearly projected into embeddings, and prepended with a learnable class token, forming a sequence of length $N+1$. Positional embeddings are added to retain spatial information. Given block output $Z^{(l)}$, the self-attention in block $l$ is defined as:  

\begin{equation}
\text{Attn}(Q,K,V) = \text{Softmax}\!\left(\frac{QK^{T}}{\sqrt{d_k}}\right)V,
\end{equation}

where $Q,K,V \in \mathbb{R}^{(N+1) \times d}$ are the query, key, and value matrices derived from $Z^{(l)}$ (with $Z^{(0)}$ being the flattened patch embeddings).  
\subsection{Token Pruning in ViTs}
The three most commonly employed token pruning frameworks that are targeted by Adversarial Efficiency Degradation Attacks on ViTs include:
\begin{itemize}
\item \textbf{ATS} (Adaptive Token Sampling) \cite{ats} ranks token significance per-block using class-token attention and samples a subset, producing a reduced sequence for subsequent blocks.
\item \textbf{AdaViT} \cite{adavit} attaches per-block decision networks that activate/deactivate tokens, heads, and block components via differentiable Gumbel-Softmax, learning compact execution paths.
\item \textbf{A-ViT} \cite{avit} equips each token with a scalar halting score that accumulates across layers; when a token exceeds a threshold, it \emph{halts} and is excluded from inference.
\end{itemize}
These mechanisms reduce \emph{average-case} FLOPs while maintaining accuracy near the static ViT baseline.
\subsection{Adversarial Efficiency Degradation Attacks: Definition}
Traditional adversarial attacks exploit neural networks by introducing imperceptible perturbations to the input, causing misclassification. Formally, a clean input $x$ is modified into an adversarial input $x_e = x + \delta$, where $\delta$ denotes the perturbation. The objective may be to force the model to predict a specific label $l_0$, i.e., $f(x_e) = l_0$, or any incorrect label different from the ground truth.  

In contrast, \textit{Adversarial Efficiency Degradation Attacks} apply perturbations with a different goal: instead of altering prediction correctness, they aim to degrade system performance by increasing resource usage. Let $N(\cdot)$ denote a neural network, $x \in \mathcal{X}$ an input, and $O$ an inference optimization method. The computational cost is represented as $C(N,O,x)$, measured in terms of energy, latency, or FLOPs. The adversary seeks a perturbation $\delta$ that maximizes computational cost while remaining imperceptible:  

\begin{equation}
\begin{aligned}
\max_{\delta} \quad & C(N,O,x+\delta) \\
\text{s.t.} \quad & \|\delta\| \leq \epsilon, \quad x+\delta \in \mathcal{X},
\end{aligned}
\end{equation}

where $\epsilon$ bounds the perturbation magnitude. The objective is to inflate computational overhead without violating input constraints.  
A typical strategy involves employing the PGD attack while adapting the loss function to align with the behavior of the specific optimization method under consideration. Additionally, as a secondary objective, the adversary seeks to enhance the stealthiness of the attack by preserving the model’s original prediction.
\subsection{Adversarial Efficiency Degradation Attacks: Past Works}
Only recently has research begun to explore adversarial attacks specifically aimed at efficient neural networks. One of the earliest attempts, ILFO \cite{haque2020ilfo}, targeted two CNN-based input-adaptive architectures—SkipNet \cite{wang2018skipnet} and SACT \cite{figurnov2017spatially}—using image-specific perturbations. However, ILFO suffered from instability in dynamic-width networks because of its sensitivity to gating mechanisms. To overcome this limitation, GradAuto \cite{pan2022gradauto} proposed a multi-objective optimization scheme that balanced gradient sensitivity across gates, ensuring more consistent perturbations. It further introduced directional gradient optimization to reduce interference between active and inactive gates, selectively deepening execution paths.
% Building on these ideas, GradMDM \cite{gradmdm} advanced FLOPs inflation by incorporating a power-oriented loss function, which favored expensive execution routes over uniform gate activations.
Another line of work, DeepSloth \cite{hong2020panda}, specifically targeted early-exit models, suppressing confidence at intermediate classifiers to block early termination and enforce full-depth execution. Similarly, GradAuto proved effective in attacking networks with both dynamic width and depth, while NICGSlowDown and TransSlowDown \cite{chen2022transslowdown, chen2022nicgslowdown} extended adversarial efficiency degradation attacks to sequence-based tasks such as neural image captioning and machine translation.
However, in this work, we concentrate on Adversarial Efficiency Degradation Attacks targeting ViTs, with particular emphasis on the widely adopted optimization strategy of token pruning.

\section{Adversarial Efficiency Degradation Attacks}
In this section, we address the research questions (RQ1 \& RQ2) outlined in Section~\ref{sec:1}. Our goal is to clarify why input-adaptive inference is vulnerable and, to unpack the mechanisms behind efficiency-targeted attacks.

\subsection{Efficiency Vulnerability of Dynamic Inference (RQ1)}
\label{subsec:efficiency-attack-avit}

% --- RQ1 statement (your text, lightly cleaned) ---
In this subsection we aim to answer \textbf{RQ1}. We explain how an input-adaptive optimization with a compute policy that reduces computation can be exploited to \emph{increase} or even \emph{reverse} the intended efficiency.

Input-adaptive methods implement a \emph{compute policy} that maps an image to a pattern of kept/removed computation. Let $\pi(x)$ denote this policy (e.g., token masks or halting indices), and let $C(x)$ be the resulting cost (FLOPs/latency/energy). A model is \emph{efficiency-vulnerable} if there exist perceptually small changes $\delta$ (e.g., $\lVert \delta \rVert_p \le \varepsilon$) such that $C(x+\delta)$ is driven toward a worst-case regime (e.g., $C(x+\delta) \approx C_{\text{dense}}$) while the task output is preserved or acceptably degraded. Mechanistically, this occurs whenever $\pi(x)=\Phi\!\big(g(x)\big)$ depends on \emph{image-conditioned, differentiable statistics} $g(x)$ (attention maps, mask logits, halting scores, confidence margins, expert gates, etc.). By slightly shifting $g(x)$ into ``keep-more / exit-later / activate-wider'' regions, an adversary can consistently increase compute. Thus, any input-adaptive optimization that ties computation to smooth, content-driven signals inherits the same class of vulnerability: the \emph{compute path} becomes as steerable from pixel space as the \emph{decision boundary} is in traditional adversarial attacks.

% --- Concise A-ViT example to append ---
\textbf{Illustrative example: A-ViT.}
A-ViT instantiates $\pi(x)$ via per-token \emph{halting} scores that accumulate across depth. For block-$l$ embeddings $Z^l$ and token $j$, a lightweight head $H(\cdot)$ produces
\begin{equation}
h^l_j \;=\; H(Z^l_j)\;=\;\sigma\!\big(\gamma\,Z^l_{j,0}+\beta\big)\in[0,1],
\end{equation}
and the token halts at the first index $I_j$ whose partial sum crosses a fixed threshold:
\begin{equation}
I_j \;=\; \arg\min_{n\le L}\;\sum_{l=1}^{n} h^l_j \;\ge\; 1-\tau.
\end{equation}
On ``easy'' inputs with large homogeneous regions, many tokens quickly accumulate mass and halt early, so $\pi(x)$ keeps few tokens active in deep layers and $C(x)$ is low. On cluttered inputs, partial sums grow more slowly, more tokens survive longer, and $C(x)$ rises. Because each $h^l_j$ is a smooth function of $Z^l_j$ and hence of the pixels, small input changes can smoothly shift the halting trajectories $\{\sum_{l\le n}h^l_j\}$, delaying halts across tokens without changing the predicted label. In turn, $\pi(x)$ flips from ``early-stop for many tokens'' to ``keep-most tokens deep,'' pushing $C(x)$ back toward the dense baseline. 
% This mechanism exemplifies how a dynamic, input-adaptive policy—here, token halting—exposes a controllable compute path and thus an efficiency-focused attack surface.

\subsection{Mechanisms of Efficiency-Targeted Attacks (RQ2)}
\label{sec:mechanisms-rq2}

\subsubsection{DeSparsify}
\label{subsec:desparsify}
DeSparsify \cite{desparsify} is a \emph{pixel-level} adversarial perturbation that maximizes the model’s compute while preserving the original label. It optimizes an input-dependent perturbation $\delta$ (with $\|\delta\|_p\le\varepsilon$) using a loss that directly \emph{thwarts} the sparsification rule of the target method, combined with a classification-preservation term. Formally, with compute-oriented loss $\mathcal{L}_{\text{atk}}$ and a label-preserving term $\mathcal{L}_{\text{cls}}$, DeSparsify performs projected (sign) gradient steps
\[
\delta \leftarrow \Pi_{\|\delta\|\le \varepsilon}\big(\delta + \alpha\,\mathrm{sgn}(\nabla_\delta [\mathcal{L}_{\text{atk}} + \lambda\,\mathcal{L}_{\text{cls}}])\big),
\]
and instantiates $\mathcal{L}_{\text{atk}}$ to flip the specific dynamic policy from ``prune'' to ``keep'' for each token-pruning mechanism. The paper demonstrates single-image, class-universal, and universal variants and evaluates on ATS, AdaViT, and A-ViT, showing large increases in FLOPs, memory use, and latency while maintaining the clean prediction \cite{desparsify}.

\textbf{Against A-ViT.}
A-ViT halts a token once the cumulative halting score exceeds $1-\tau$. With $h^l_j=\sigma(\gamma Z^l_{j,0}+\beta)$ and $I_j=\arg\min_{n\le L}\sum_{l=1}^n h^l_j\ge 1-\tau$, DeSparsify lowers partial sums across depth so tokens \emph{fail to halt}. A convenient surrogate penalizes all pre-halt partial sums toward $0$:
\[
\mathcal{L}_{\text{A-ViT}}
=\frac{1}{N}\sum_{j=1}^{N}\frac{1}{L}\sum_{n=1}^{L}\mathbb{1}_{I_j<n}\,\ell_{\text{MSE}}\!\Big(\sum_{l=1}^{n}h^l_j,\;0\Big),
\]
which drives $I_j\!\to\!L$ and restores near-dense compute \cite{desparsify}. 

\textbf{Against AdaViT.}
AdaViT’s decision networks emit per-block binary masks over blocks/heads/patches via Gumbel-Softmax relaxations $M_l=\big(\mathrm{GS}(m^b_l),\mathrm{GS}(m^h_l),\mathrm{GS}(m^p_l)\big)$. It pushes these toward ``activate'' so nothing is pruned:
% --- AdaViT attack loss split into three compact parts ---

% (1) Block-mask term (keeps both submodules active)
\begin{equation}
\mathcal{L}^{(l)}_{\text{block}}
\;=\;
\frac{1}{2}\sum_{b=1}^{2}\ell_{\text{MSE}}\!\big(M^{b}_{l},\mathbf{1}\big).
\end{equation}

% (2) Head-mask term (only if MSA block is active)
\begin{equation}
\mathcal{L}^{(l)}_{\text{head}}
\;=\;
\mathbf{1}\!\{M^{b}_{l,\mathrm{MSA}}=1\}\;
\frac{1}{H}\sum_{h=1}^{H}\ell_{\text{MSE}}\!\big(M^{h}_{l},\mathbf{1}\big).
\end{equation}

% (3) Patch-mask term (keeps all tokens)
\begin{equation}
\mathcal{L}^{(l)}_{\text{patch}}
\;=\;
\frac{1}{N}\sum_{p=1}^{N}\ell_{\text{MSE}}\!\big(M^{p}_{l},\mathbf{1}\big).
\end{equation}

% Aggregate per-layer and across layers (fits in one line)
\begin{equation}
\mathcal{L}^{(l)}_{\text{AdaViT}}
=\mathcal{L}^{(l)}_{\text{block}}
+\mathcal{L}^{(l)}_{\text{head}}
+\mathcal{L}^{(l)}_{\text{patch}},
\qquad
\mathcal{L}_{\text{AdaViT}}
=\frac{1}{L}\sum_{l=1}^{L}\mathcal{L}^{(l)}_{\text{AdaViT}}.
\end{equation}

conditioning head activation on the MSA decision \cite{desparsify}. 

\textbf{Against ATS.}
ATS samples survivors from significance scores $S_j\propto A_{1,j}\,\|V_j\|$. To maximize retained tokens, DeSparsify \emph{flattens} the score distribution (high entropy), e.g., by minimizing a divergence to the uniform target, which makes inverse-CDF samples mostly distinct and prevents pruning. 

\medskip

\subsubsection{SlowFormer}
\label{subsec:slowformer}
Unlike DeSparsify’s \emph{pixel-level} perturbations crafted per image (or per class), \textbf{SlowFormer} \cite{slowformer} is a \emph{universal adversarial patch}: a small learned patch $p$ (fixed location/size) trained once on a dataset and pasted onto \emph{all} test images without per-sample optimization. The patch is updated by gradient ascent on a compute-maximization loss tailored to the target method, while the backbone weights remain frozen; pixels are projected back to the valid range each step. This patch-based, input-agnostic form is argued to be more practical for real-world attacks than per-image perturbations, yet it can still restore compute close to the non-efficient baseline across A-ViT, ATS, and AdaViT. 

\textbf{Against A-ViT.}
SlowFormer reuses A-ViT’s own training objectives but \emph{negates} them to delay halting and increase compute. If $L_{\text{task}}$ is the classification loss, $L_{\text{ponder}}$ promotes early halts, and $L_{\text{distr}}$ regularizes halting-score statistics, the patch maximizes compute by minimizing 
\[
-\big(\alpha_d L_{\text{distr}}+\alpha_p L_{\text{ponder}}\big),
\]
optionally adding $+\,L_{\text{task}}$ to preserve accuracy. Training optimizes only patch pixels; at test time, the learned patch reliably recovers pruned tokens and raises GFLOPs toward the dense model \cite{slowformer}. 

\textbf{Against AdaViT.}
AdaViT uses a computation-target term and a usage loss to keep masks sparse. SlowFormer sets the computation-target $\gamma\!=\!0$ and \emph{negates} the usage loss so the patch increases the keep probabilities of patches/heads/blocks; a cross-entropy term can be included or omitted depending on whether accuracy preservation is desired. Compute increases are smaller than A-ViT’s upper bound but still significant \cite{slowformer}. 

\textbf{Against ATS.}
Since ATS selects tokens via inverse-transform sampling from $S$ (class attention $\times$ value norm), SlowFormer directly \emph{uniformizes} the class token’s attention over patches to maximize unique survivors:
\[
\mathcal{L}_{\text{ATS}}=\sum_{i=2}^{N}\big\|A_{1,i}-\tfrac{1}{N}\big\|_2^2,
\]
summing over heads and layers as needed. The learned patch induces near-uniform attention, preventing pruning and increasing FLOPs, latency, and energy \cite{slowformer}. 

% \medskip
% \noindent\textbf{Summary.} DeSparsify and SlowFormer share the same principle: instantiate a differentiable surrogate for each dynamic policy and optimize the input (pixels or a universal patch) so the policy outputs ``keep/activate'' rather than ``prune/halt.'' DeSparsify is per-image (or per-class) and can preserve the original label explicitly; SlowFormer is a single transferable patch trained once and applied everywhere. Both reliably steer the compute path toward the worst-case, reversing the intended savings of token-sparsifying ViTs. 
\section{Defenses Against Efficiency-Targeted Attacks}
\label{sec:defenses}
To date, there is no dedicated, comprehensive defense specifically against to the two attacks studied here. Both works, however, outline some strategies to secure token-pruning ViTs against efficiency-focused adversaries. 

For \textbf{SlowFormer} \cite{slowformer}, standard adversarial training is adapted from per-image perturbations to universal patches. Instead of crafting a fresh patch per sample, the training loop maintains a small pool of adversarial patches and, at each iteration, draws one uniformly at random to paste on the input while optimizing the model’s original task loss. To keep the patch pool aligned with the evolving model, training is periodically paused (e.g., every 20\% of an epoch) to optimize a new patch for a limited number of steps and add it to the pool. This procedure preserves the practicality of patch-based training while containing compute, yet yields a model that is measurably more robust to universal efficiency degradation attacks. 

\textbf{DeSparsify} \cite{desparsify} suggests two ways to increase robustness. First, increasing the parameterization of the token-sparsification mechanism (e.g., a learned decision network as in AdaViT versus a single-neuron halting signal as in A-ViT) tends to raise robustness, whereas mechanisms tied to non-purpose-optimized signals (e.g., raw attention scores as in ATS) can be more brittle. Second, enforcing a per-block \emph{upper bound} on active tokens---estimated from the average number of tokens on a holdout set---constrains worst-case compute under attack while preserving most of the clean-image savings. When this cap is exceeded, tokens can be dropped either at random or via a simple confidence heuristic.
%----------------------- SETUP SUMMARY TABLE (simplified) -----------------------%
\section{Experimental Results}
%\subsection{Setup}
%---------------- SETUP SUMMARY (one-column, concise, vertical) ----------------%
\begin{table}[t]
\centering
\caption{Setup summary for \textit{SlowFormer} (2023) and \textit{DeSparsify} (2024).}
\label{tab:setup-summary}
\footnotesize
\renewcommand{\arraystretch}{1.05}
\setlength{\tabcolsep}{5pt}
\begin{tabularx}{\columnwidth}{l X}
\toprule
\textbf{Attack} & \textbf{Key setup details} \\
\midrule
\textbf{SlowFormer} &
\textit{Backbones/TP:} ViT (Tiny/Small/Base) with A-ViT, ATS, AdaViT. \newline
\textit{Datasets:} ImageNet-1K (main), CIFAR-10 (also). \newline
\textit{Attack/Budget:} Universal square patch (e.g., $64{\times}64$), fixed location; optimized (AdamW). \newline
\textit{Metrics:} GFLOPs, Top-1 accuracy, attack success (\emph{no direct power}). \newline
\textit{Placement:} Standard ATS/A-ViT/AdaViT layer positions (ATS typically mid blocks). \newline
\textit{Impl./HW:} PyTorch; RTX~3090-class GPUs. \\
\midrule
\textbf{DeSparsify} &
\textit{Backbones/TP:} DeiT (Tiny/Small), T2T-ViT-19 with ATS, AdaViT, A-ViT. \newline
\textit{Datasets:} ImageNet (val), CIFAR-10 (val). \newline
\textit{Attack/Budget:} Iterative $\ell_\infty$ perturbation (e.g., $\epsilon{=}16/255$), cosine schedule. \newline
\textit{Metrics:} TUR (token utilization ratio), GFLOPs, accuracy; \emph{HW signals} (e.g., memory/throughput). \newline
\textit{Placement:} ATS later blocks (e.g., 4–12), AdaViT earlier decisions; A-ViT early halting. \newline
\textit{Impl./HW:} PyTorch; RTX~3090-class GPUs. \\
\bottomrule
\end{tabularx}
\vspace{-2mm}
\end{table}
%----------------------- END TABLE -----------------------%

%----------------------- BRIEF SETUP SYNOPSIS (simplified) -----------------------%

Table~\ref{tab:setup-summary} depicts the setup for both attacks. \textit{SlowFormer} builds a single adversarial patch that generally pushes adaptive ViTs (A-ViT, ATS, AdaViT) to keep more tokens and therefore spend more FLOPs, it reports compute (GFLOPs) and accuracy but does not measure power directly. \textit{DeSparsify} uses an $\ell_\infty$-bounded perturbation to raise the token count (TUR), inflate GFLOPs, and also reports hardware-side effects like memory and throughput. Both evaluate on ImageNet/CIFAR with similar ViT-family backbones and same token pruning frameworks.
% Requires: \usepackage{multirow}
%==== ViT/DeiT-Small: Framework × Attack with Accuracy Loss and Attack Success ====%
\subsection{Comparsion between patch and single efficiency adversarial attack}
\begin{table}[t]
\centering
\caption{Attack results of DeSparsify (on DeiT-S) and SlowFormer (on ViT-S)  for token pruning frameworks, A-ViT, ATS and AdaViT.}
\label{tab:vitS_framework_attack_accLoss_AS}
\footnotesize
\scriptsize
\setlength{\tabcolsep}{2pt}
\renewcommand{\arraystretch}{1.08}
\begin{tabular}{l l c c c}
\toprule
\textbf{Token-pruning framework} & \textbf{Attack type} &
\shortstack{GFLOPs\\(attack)} &
\shortstack{Accuracy\\loss (\%)} &
\shortstack{Attack\\Success (\%)} \\
\midrule
\textbf{ViT-Small (baseline)} &  & \textbf{4.60} & N/A & N/A \\
\midrule
\multirow{3}{*}{\textbf{A-ViT}}
  & \quad No attack            & 3.70 & 0.0  & -- \\
  & \quad SlowFormer (patch)   & \textbf{4.60} & 76.5 & 100.0 \\
  & \quad DeSparsify (single)  & 4.60 & 0.1  & 100.0 \\
\midrule
\multirow{3}{*}{\textbf{ATS}}
  & \quad No attack            & 3.10 & 0.0  & -- \\
  & \quad SlowFormer (patch)   & 4.00 & 78.2 & 60.0 \\
  & \quad DeSparsify (single)  & 4.20 & 1.2  & 73.3 \\
\midrule
\multirow{3}{*}{\textbf{AdaViT}}
  & \quad No attack            & 2.25 & 0.0  & -- \\
  & \quad SlowFormer (patch)   & 3.20 & 76.9 & 40.4 \\
  & \quad DeSparsify (single)  & 3.27 & 0.2  & 43.4 \\
\bottomrule
\end{tabular}
\vspace{-2mm}
\end{table}
In Table \ref{tab:vitS_framework_attack_accLoss_AS}, we compare two efficiency-focused attacks on token-pruned vision transformers: \emph{SlowFormer}, which uses a universal patch (a single, image-agnostic square patch optimized once and pasted at a fixed location on every input), and \emph{DeSparsify (single)}, which applies a per-image additive perturbation (tailored noise, bounded in $\ell_\infty$, added to the whole image). For each token-pruning framework (A-ViT, ATS, AdaViT), the table reports the resulting GFLOPs, the \emph{accuracy loss} (drop relative to the framework’s own “No attack” accuracy), and the \emph{Attack Success (AS)}, computed as
\[
\mathrm{AS}=\frac{F_{\text{attack}}-F_{\min}}{F_{\max}-F_{\min}}\times 100\%,
\]
where $F_{\min}$ is the “No attack” GFLOPs for that framework and $F_{\max}$ is the dense ViT-Small GFLOPs. We show ViT-Small results for SlowFormer and DeiT-Small results for DeSparsify; because DeiT-Small and ViT-Small share the same ViT-S architecture, the GFLOPs and accuracy-loss comparisons are directly meaningful across the two sources, and we present them consistently across the three token-pruning frameworks: A-ViT, ATS and AdaViT. Overall, both attacks increase GFLOPs by pushing the pruning policy to retain more tokens, thereby reducing efficiency and causing measurable accuracy loss. In many cases, the \emph{single} (per-instance) attack attains slightly higher attack success rate than the universal patch, plausibly because it adapts to the image content and the current pruning decisions, whereas a universal patch must work across diverse inputs and its effectiveness is sensitive to fixed design choices such as patch size and placement. 
Table \ref{tab:vitS_framework_attack_accLoss_AS} can not only increase computation but also reduce model accuracy, depending on the attacker’s goals. A drop in accuracy may be a useful side effect—much like in standard adversarial attacks, but it can also make the attack easier to detect. These attacks can be optimized either to target computational cost while either preserving or degrading task performance by incorporating a task loss into the optimization objective.
% Requires: \usepackage{multirow}
%============= Combined Defense Results: DeSparsify (single) + SlowFormer (adv training) =============%
% Requires: \usepackage{multirow}
%============= Combined Defense Results (GFLOPs + AS only) =============%
% Requires: \usepackage{multirow}
%============= Combined Defense Results (GFLOPs + AS for both papers) =============%
\begin{table}[t]
\centering
\caption{Defense effectiveness on Adversarial Efficiency Degradation attacks. 
DeSparsify (DeiT-S) uses a confidence-based rule; 
SlowFormer (ViT-S, A-ViT) uses adversarial training. 
Entries report GFLOPs and Attack Success (AS) \emph{before} and \emph{after} defense, showing reduced compute inflation.}
\label{tab:combined-defenses-concise-noacc}
\footnotesize
\setlength{\tabcolsep}{5pt}
\renewcommand{\arraystretch}{1.08}
\begin{tabular}{l l l c c}
\toprule
\textbf{Model} &
\shortstack{\textbf{Token-pruning}\\\textbf{framework}} &
\textbf{Setting} &
\shortstack{\textbf{GFLOPs}} &
\shortstack{\textbf{AS}\\\textbf{(\%)}} \\
\midrule
\multirow{6}{*}{\textbf{DeSparsify}}
  & ATS     & No defense   & 4.20 & 73.5 \\
  & ATS     & With defense & 3.17 & 5.3  \\
  & Ada-ViT & No defense   & 3.27 & 43.2 \\
  & Ada-ViT & With defense & 2.36 & 4.3  \\
  & A-ViT   & No defense   & 4.60 & 100.0 \\
  & A-ViT   & With defense & 3.95 & 36.9 \\
\midrule
\multirow{3}{*}{\textbf{SlowFormer}}
  & A-ViT & No attack     & 0.87 & --   \\
  & A-ViT & No defense    & 1.26 & 100  \\
  & A-ViT & With defense  & 1.01 & 34   \\
\bottomrule
\end{tabular}
\vspace{-2mm}
\end{table}

% Notes:
% • DeSparsify rows show the reported GFLOPs for adversarial images under the single-image attack,
%   comparing “No defense” vs. “Confidence” defense. Accuracy and AS were not reported in that table.
% • SlowFormer rows (A-ViT, ViT-S) include GFLOPs, Top-1 accuracy, and AS with and without adversarial training.

%========================================================================================%

%====================================================================================%

%===============================================================================%

% ---------------------------------------------------------

% ---------------------------------------------------------
% Narrative tie-in (optional paragraph)
% % ---------------------------------------------------------

% \paragraph*{Overall.}
% Across token-adaptive ViTs, both universal-patch and per-image PGD attacks consistently suppress adaptivity and drive worst-case compute. Masking mechanisms primarily reveal the effect through GFLOPs/TUR, while removal mechanisms reflect corresponding hardware penalties. Defenses that restore selection/halting decisiveness can limit compute inflation while preserving a significant fraction of clean-image efficiency.

% ================================
% End Results
% ================================
\subsection{Defenses against adversarial efficiency degradation attacks.} 
Table~\ref{tab:combined-defenses-concise-noacc} summarizes the defense results reported in the two papers. For \emph{DeSparsify} (evaluated on DeiT-S), we present the authors’ \emph{confidence-based} defense, which adjusts token-retention decisions using a confidence signal to counter de-sparsification. For \emph{SlowFormer} (evaluated on ViT-S with A-ViT), we show adversarial training, where the model is trained against a pool of universal patches to improve robustness. The table reports GFLOPs and Attack Success rate (AS) before and after applying each defense. DeSparsify’s confidence-based defense reduces AS substantially across all three token-pruning frameworks: on \textbf{ATS}, AS drops from \(\mathbf{73.5\%}\) (no defense) to \(\mathbf{5.3\%}\) (with defense); on \textbf{AdaViT}, from \(\mathbf{43.2\%}\) to \(\mathbf{4.3\%}\); and on \textbf{A\,-\,ViT}, from \(\mathbf{100\%}\) to \(\mathbf{36.9\%}\). For SlowFormer’s universal-patch attack on \textbf{A\,-\,ViT}, adversarial training lowers AS from \(\mathbf{100\%}\) to \(\mathbf{34\%}\), while also pulling GFLOPs closer to the clean operating point. These results indicate that, although neither defense fully restores the pre-attack efficiency, both measurably curb compute inflation and therefore make token-sparsified inference more robust. 
\section{Discussion}
In this section, we examine the hardware implications of these attacks and the limitations of efficiency adversarial methods. We also analyze current defensive strategies, highlight open challenges, and outline directions for future work.

\textbf{Effects on hardware:}
Efficiency-targeted attacks are particularly harmful in constrained or latency-sensitive settings such as edge devices, mobile platforms, and real-time systems. By forcing token-pruning mechanisms to retain more tokens, these attacks inflate core resource demands—\emph{memory footprint}, \emph{energy consumption}, and \emph{execution time}. For example, the single-image variant of \emph{DeSparsify} is reported to raise memory usage by roughly \textbf{37\%}, increase energy consumption by about \textbf{72\%}, and alter throughput by around \textbf{8\%} relative to clean inputs (i.e., fewer images processed per unit time under attack). In addition, as noted by the authors of ATS~\cite{ats}, activating ATS introduces a modest overhead from its sampling I/O, which manifests as extra runtime compared with the vanilla (non-pruning) model. 

\noindent\textbf{Limitations of Adversarial Efficiency Degradation Attacks:}
We highlight three key limitations observed in current methods: (1) \emph{Dependence on internal mechanisms}—most attacks assume knowledge of the token-pruning policy (e.g., scoring signals, halting rules, retention caps) to steer decisions toward higher compute, which reduces effectiveness in \emph{black-box} settings where such internals are hidden or randomized; (2) \emph{Limited transferability}—adversarial examples crafted for one token-pruning framework (e.g., ATS) often transfer poorly to others (e.g., A-ViT or AdaViT), implying that separate perturbations may be needed per framework and increasing attack cost and complexity; and (3) \emph{Narrow applicability}—these attacks primarily target \emph{input-adaptive} inference (token pruning/early halting) and have little to no effect on \emph{static} efficiency techniques such as model pruning or architecture-level slimming, where computation is fixed at inference time.

\noindent\textbf{Issues in current defenses:} Existing countermeasures against adversarial efficiency degradation attacks exhibit several gaps: (1) most defenses are tailored to a single token-sparsification mechanism, so protection is brittle when the pruning policy or backbone changes; (2) efficacy is often reported in FLOPs only, while hardware-facing outcomes (energy, memory, latency) are rarely measured in-the-loop; (3) adversarial training or confidence gating can reduce attack leverage but also erode clean-time efficiency, and this trade-off is not systematically optimized; (4) fixed thresholds or handcrafted heuristics are easy to reverse-engineer, enabling adaptive attacks, whereas injected randomness is often superficial; (5) defenses tuned for specific budgets, patch sizes, or $\ell_\infty$ radii underperform outside those regimes and transfer poorly across datasets or input scales; and (6) there are no certified bounds on “efficiency robustness” (e.g., worst-case token inflation), leaving safety-critical settings without guarantees.

% \noindent\noindent\textbf{Future directions for defenses.} Promising defense directions include: (1) budget-aware robust training that jointly preserves accuracy and constrains worst-case compute by penalizing adversarial token inflation or retention-entropy collapse; (2) stochastic, learnable gates (e.g., Gumbel/Concrete relaxations) trained with adversarial regularization to make exact gradients and attack trajectories less reliable; (3) co-training against a family of token-sparsifiers (ATS/A-ViT/AdaViT) and distilling shared invariants to strengthen cross-policy consistency; (4) runtime monitors that use token-utilization ratios, memory bandwidth, or power spikes to detect de-sparsification and trigger fallbacks (caps, threshold shifts, or backbone switching); (5) patch-aware denoising or masking that suppresses spatially localized, input-agnostic cues without harming salient content.

% \textbf{Future directions for attacks.} On the attack side, useful directions are: (1) black-box, query-efficient methods using score-only or decision-based probes to infer retention gradients without internal access; (2) cross-policy transfer by optimizing perturbations that inflate compute across multiple pruning policies and backbones simultaneously; (3) multi-objective, hardware-aware attacks that co-optimize compute, memory traffic, and power under perceptual constraints, including strategies that target I/O bottlenecks or on-chip buffers.
\noindent\textbf{Future Directions:}
Going forward, attacks should aim to \emph{transfer} across different inference optimizations and also cover settings that are still unexplored, including black-box and physical scenarios. On the defense side, the goal should be to provide \emph{guarantees} on protection against all kind of adversarial efficiency degradation attacks.

\section{Conclusion}
This survey systematizes \emph{Adversarial Efficiency Degradation Attacks} on ViTs and clarifies how input-adaptive policies expose a compute pathway that is steerable from pixel space. Using a common lens over SlowFormer and DeSparsify and three token-pruning frameworks, we show consistent \emph{GFLOPs inflation} and  use \emph{Attack Success} metric to make cross-paper comparisons concrete. We further summarize defenses that reduce token inflation (confidence-based caps) or improve robustness to universal patches (adversarial training), noting that both reduce, but do not fully remove attack leverage. 

\bibliographystyle{IEEEtran}
{\footnotesize\bibliography{refs}}
\end{document}